\documentclass[11pt]{article}
\usepackage{amsfonts,amssymb, graphicx,a4,bm}
\newcommand{\qed}{\hfill\rule{3mm}{3mm}}
\newtheorem{teorema}{Theorem}%[section]
\newtheorem{Remark}[teorema]{Remark}%[section]

\begin{document}

\def\reff#1{(\protect\ref{#1})}

\let\a=\alpha \let\b=\beta \let\ch=\chi \let\d=\delta \let\e=\varepsilon
\let\f=\varphi \let\g=\gamma \let\h=\eta    \let\k=\kappa \let\l=\lambda
\let\m=\mu \let\n=\nu \let\o=\omega    \let\p=\pi \let\ph=\varphi
\let\r=\rho \let\s=\sigma \let\t=\tau \let\th=\vartheta
\let\y=\upsilon \let\x=\xi \let\z=\zeta
\let\D=\Delta \let\F=\Phi \let\G=\Gamma \let\L=\Lambda \let\Th=\Theta
\let\O=\Omega \let\P=\Pi \let\Ps=\Psi \let\Si=\Sigma \let\X=\Xi
\let\Y=\Upsilon
%%%%%%%%%%%%%%%% EQUAZIONI CON NOMI SIMBOLICI
%%% per assegnare un nome simbolico ad una equazione basta
%%% scrivere \Eq(...) o, in \eqalignno, \eq(...) o,
%%% nelle appendici, \Eqa(...) o \eqa(...):
%%% dentro le parentesi e al posto dei ...
%%% si puo' scrivere qualsiasi commento; per avere i nomi
%%% simbolici segnati a sinistra delle formule si deve
%%% dichiarare il documento come bozza, iniziando il testo con
%%% \BOZZA. Sinonimi \Eq,\EQ.
%%% All' inizio di ogni paragrafo si devono definire il
%%% numero del paragrafo e della prima formula dichiarando
%%% \numsec=... \numfor=... (brevetto Eckmannn).

\global\newcount\numsec\global\newcount\numfor
\gdef\profonditastruttura{\dp\strutbox}
\def\senondefinito#1{\expandafter\ifx\csname#1\endcsname\relax}
\def\SIA #1,#2,#3 {\senondefinito{#1#2}
\expandafter\xdef\csname #1#2\endcsname{#3} \else
\write16{???? il simbolo #2 e' gia' stato definito !!!!} \fi}
\def\etichetta(#1){(\veroparagrafo.\veraformula)
\SIA e,#1,(\veroparagrafo.\veraformula)
 \global\advance\numfor by 1
% \write15{@def@equ(#1){\equ(#1)} \%:: ha simbolo= #1 }
 \write16{ EQ \equ(#1) ha simbolo #1 }}
\def\etichettaa(#1){(A\veroparagrafo.\veraformula)
 \SIA e,#1,(A\veroparagrafo.\veraformula)
 \global\advance\numfor by 1\write16{ EQ \equ(#1) ha simbolo #1 }}
\def\BOZZA{\def\alato(##1){
 {\vtop to \profonditastruttura{\baselineskip
 \profonditastruttura\vss
 \rlap{\kern-\hsize\kern-1.2truecm{$\scriptstyle##1$}}}}}}
\def\alato(#1){}
\def\veroparagrafo{\number\numsec}\def\veraformula{\number\numfor}
\def\Eq(#1){\eqno{\etichetta(#1)\alato(#1)}}
\def\eq(#1){\etichetta(#1)\alato(#1)}
\def\Eqa(#1){\eqno{\etichettaa(#1)\alato(#1)}}
\def\eqa(#1){\etichettaa(#1)\alato(#1)}
\def\equ(#1){\senondefinito{e#1}$\clubsuit$#1\else\csname e#1\endcsname\fi}
\let\EQ=\Eq
\def\0{\emptyset}

\def\pp{{\bm p}}\def\pt{{\tilde{\bm p}}}

%%%%%%%%%%%%%%% DEFINIZIONI LOCALI

\def\\{\noindent}
\let\io=\infty

\def\VU{{\mathbb{V}}}
\def\EE{{\mathbb{E}}}
\def\GI{{\mathbb{G}}}
\def\TT{{\mathbb{T}}}
\def\C{\mathbb{C}}
\def\CC{{\mathcal C}}
\def\II{{\mathcal I}}
\def\LL{{\cal L}}
\def\RR{{\cal R}}
\def\SS{{\cal S}}
\def\NN{{\cal N}}
\def\HH{{\cal H}}
\def\GG{{\cal G}}
\def\PP{{\cal P}}
\def\AA{{\cal A}}
\def\BB{{\cal B}}
\def\FF{{\cal F}}
\def\v{\vskip.1cm}
\def\vv{\vskip.2cm}
\def\gt{{\tilde\g}}
\def\E{{\mathcal E} }
\def\I{{\rm I}}

\def\tende#1{\vtop{\ialign{##\crcr\rightarrowfill\crcr
              \noalign{\kern-1pt\nointerlineskip}
              \hskip3.pt${\scriptstyle #1}$\hskip3.pt\crcr}}}
\def\otto{{\kern-1.truept\leftarrow\kern-5.truept\to\kern-1.truept}}
\def\arm{{}}
\font\bigfnt=cmbx10 scaled\magstep1

%%%%%%%%%%%%%%% DEFINIZIONI ROBERTO
\newcommand{\card}[1]{\left|#1\right|}
\newcommand{\und}[1]{\underline{#1}}
\def\1{\rlap{\mbox{\small\rm 1}}\kern.15em 1}
\def\ind#1{\1_{\{#1\}}}
\def\bydef{:=}
\def\defby{=:}
\def\buildd#1#2{\mathrel{\mathop{\kern 0pt#1}\limits_{#2}}}
\def\card#1{\left|#1\right|}
\def\proof{\noindent{\bf Proof. }}
\def\qed{ \square}
\def\trp{\mathbb{T}}
\def\trt{\mathcal{T}}
\def\Z{\mathbb{Z}}
\def\be{\begin{equation}}
\def\ee{\end{equation}}
\def\bea{\begin{eqnarray}}
\def\eea{\end{eqnarray}}

\title {Absence of phase transitions  in a class of  integer spin systems %via convergent polymer expansions
}

\author{Thiago Morais  and  Aldo Procacci \\\footnotesize{Dep. Matem\'atica-ICEx, UFMG, CP 702
Belo Horizonte - MG, 30161-970 Brazil}\\\footnotesize{ email:~  aldo@mat.ufmg.br}
}
\date{}
\maketitle
\begin{abstract}
We exhibit  a  class of integer spin systems whose free energy can
be written in term of an absolutely convergent series at any
temperature. This class includes spin systems  on $\Z^d$
interacting through infinite range pair potential polynomially
decaying at large distances $r$  at a rate $1/r^{d+\e}$ with
$\e>0$. It also contains the Blume-Emery-Griffiths model in the
disordered phase at large values of the crystal field.

\end{abstract}

\numfor=1
\numsec=1

\section{The class of spin systems: definitions and results}

\numsec=1
\numfor=1

\def\sg {{\rm supp} \,g}
\def\sG {{\rm supp} \,G}

Let $\VU$ be a countable set.  We define
an integer  spin system on $\VU$ by
supposing that in each site $x\in \VU$ there is a random  variable $\s_x$ (the spin at $x$) taking values in the
set $\{0,\pm 1, \pm 2,\dots, \pm N\}$ where $N$ is an integer.
For $U\subset \VU$, a spin configuration $\s_U$  in  $U$ is a function
$\s_U: U\to \{0,\pm 1, \pm 2,\dots, \pm N\}: x\mapsto \s_x$.
We denote by $\Si_U$ the set of all spin configurations in $U$.
Given a spin configuration $\s_\L$ in the finite ``volume'' $\L\subset \VU$,
the Hamiltonian of the system at volume $\L$ (with free boundary conditions) is defined as
$$
H_\L(\o)=\sum_{\{x,y\}\subset \L}V(x,y, \s_x,\s_y)+ D\sum_{x\in \L}\s_x^2
\Eq(Ham)
$$
Typically, the set  $\VU$ is the unit
cubic lattice $\mathbb{Z}^d$,  but theorem 1 below holds for any countable set $\VU$ regardless of its structure.
The assumptions on the pair potential $V(x,y, \s_x,\s_y)$   in the Hamiltonian \equ(Ham) are the following.
\begin{enumerate}
\item[{\bf A}.] For any pairs $x,y$ such that $\s_x\s_y=0$ we have
$$
V(x,y, \s_x,\s_y)=0  ~~~~~~~~~~~~~~~~~\Eq(hyp0)
$$
\item[{\bf B}.] There exists a positive function $J(x,y)$ and a positive number $J$ such that
$$
|V(x,y, \s_x,\s_y)|\le J(x,y),~~~~~~~~~~~~~~~~~~~~\forall \,(\{x,y\},\s_x,\s_y)\Eq(hyp1)
$$
and
$$
\sup_{x\in\VU}\sum_{y\in \VU\atop y\neq x}J(x,y)=2J
~~~~~~~~~~~~~\Eq(hyp2)
$$
\end{enumerate}
Note that assumption \equ(hyp2) includes infinite range interactions polynomially decaying
in a summable way. Note also that, when $D>J$, the  lowest-energy state (i.e. the
spin configuration $\s^0_\L\in \Si_\L$ for which $H_\L(\s_\L)$ attains its minimum) is the spin configuration $\s^0_\L=0$, i.e.  all sites have  spin equal to zero. The region $D>J$ is usually called the disordered phase and the parameter $D$ is called the crystal field.

The partition function of the system in the volume $\L$, at
inverse temperature $\b$ is given by
$$
Z_\L(\b)=\sum_{\s_\L\in \Si_\L}e^{-\b H (\s_\L)}\Eq(parttt)
$$
and the  free energy at finite volume $\L$ is given by
$$
f_\L(\b)={1\over |\L|}\ln Z_\L(\b)\Eq(free)
$$
where here and elsewhere in the paper $|\L|$ is the number of elements of the finite set $\L$.

Our results on the system above can be summarized by  the following theorem.

\vv
\\{\bf Theorem 1}. {\it Consider the spin system with Hamiltonian \equ(Ham). Under the assumptions \equ(hyp0)-\equ(hyp2),
there exist  positive numbers
$D_c$,  $\b_1$ and $\b_2$  with $D_c>J$ and
$\b_1<\b_2$ such that
the free energy $f_\L(\b)$ defined in \equ(free) can be written in terms of an absolutely
convergent series  uniformly bounded in $\L$  either if
 $\b\in [0,\b_1]$ or   $\b\in[\b_2,\infty)$ and $D>J$. Moreover  if $D\ge D_c$  then
 $f_\L(\b)$ converges absolutely for  all $\b\in [0,\infty)$.
}
\vv

We remind the reader that the standard polymer expansion has been shown to be absolutely convergent
at sufficiently high temperatures  or low density for wide class of spin systems,
either using Kirkwood-Salzburg  equations  \cite{GM, GMR} or by bounding
directly the coefficients of the expansion  \cite{PS}.
Moreover it has been shown in  \cite{I} that analyticity of a similar class of spin systems
at high temperatures or low densities can be inferred from the  Dobrushin uniqueness theorem.

On the other hand it is well known that for a wide class of
spin systems a unique zero-temperature
ground state implies
a unique Gibbs state at sufficiently low  temperatures
and hence absence of phase transitions (see e.g. \cite{LM} and references therein).
Analyticity at low-temperature via Pirogov-Sinai theory
has been established in \cite{Z} for  spin systems interacting via  finite-range potentials
  and has been extended in \cite{P}
to infinite-range potentials.

Here we present a class of spin systems, depending on a positive parameter $D$,
whose free energy is analytic simultaneously at
high and low temperatures if $D>J$ or at any temperature, if $D>D_c$.
It is worth remarking that our class of spin systems  also includes the
Blume-Emery-Griffiths (BEG)   model \cite{BEG} on $\Z^d$. To recover the BEG model just choose
$\VU=\Z^d$, $N=1$  and
$$
V(x,y, \s_x,\s_y)= -V_{\{x,y\}}\s_x\s_y+ K_{\{x,y\}}\s^2_x\s^2_y
$$
with $V_{\{x,y\}}=V>0$, $K_{\{x,y\}}=K\in \mathbb{R}$ if $x,y$ are nearest neighbors and
$V_{\{x,y\}}=K_{\{x,y\}}=0$ otherwise.

Specifically on the BEG model, there are some rigorous results \cite{LR, CGLR}   concerning  the
uniqueness of the  equilibrium state in the regime where the ferromagnetic interaction $V$ is sufficiently strong.
Our results  imply that  no phase transition can occur in the BEG model when the crystal field is sufficiently large.

%%%%%%%%%%%%%%%%%%%%%%%%%%%%%%%%%%%%%%%%%%%%%%%%%%%%%%%%%%%%%%%%%%%%%%%%%%%%%%%%%%%%%%%%%%%%%%%%%%%%%%%%%%%%%%%%%%

\numfor=1\numsec=2

\section{Proof of theorem 1}

We will prove the theorem by performing a high temperature polymer expansion on the partition function of the system.
We will see below that such an
expansion, when applied to  a spin system described by the Hamiltonian \equ(Ham), converges
also in the region of low temperatures (large $\b$) as soon as $D>J$
and the gap between the analytic high temperature phase
and the analytic low-temperature phase depends on the strength of the crystal field $D$. When $D$ is sufficiently
large, this gap disappears.

\\We begins by rewriting the partition function \equ(parttt)
of the system in the volume $\L\subset \VU$ as follow.

$$
Z_\L(\b)=\sum_{\s_\L\in \Si_\L}e^{-\b H (\s_\L)}= \sum_{\s_\L\in \Si_\L} e^{-\b D\sum_{x\in \L} \s_x^2}~~e^{-\b\sum_{\{x,y\}\subset \L}V(x,y, \s_x,\s_y)}
\Eq(high)
$$
We then expand the exponential in l.h.s. of \equ(high) as follows.
\vskip.2cm
$$
e^{-\b\sum_{\{x,y\}\subset \L}V(x,y, \s_x,\s_y)}~=~
\prod_{\{x,y\}\subset\L} [e^{-\b\sum_{\{x,y\}\subset\L}V(x,y, \s_x,\s_y)}-1+1]~=~
$$
$$
~=~ \sum_{s=1}^{|\L|}\sum_{\{R_1 ,\dots ,R_s\}\in\pi_s(\L)}\r(R_1,\s_{R_1})\cdots\r(R_s,\s_{R_s})
$$
Here $\pi_s(\L)$ denotes the set of all partitions of $\L$ in $s$ non empty subsets, and \vskip.3cm
$$
\r(R,\s_{R})~=~\cases{ 1 &if $|R|~=~1$\cr\cr \sum_{g\in G_R}\prod_{\{x,y\}\in
E_g}[e^{-\b V(x,y, \s_x,\s_y)}-1] &if $|R|\geq 2$}
$$
with $G_R$ being the set of all connected
graphs  with vertex set $R$ ($E_g$ denotes the edge set of a graph $g\in G_R$). Thus \equ(high) becomes, \vskip.5cm
$$
Z_{\L}(\b)  ~=~  \sum_{{\s}_{\L}\in \Si_\L} e^{-\b D\sum_{x\in \L} \s_x^2}
\sum_{s=1}^{|\L|}\sum_{R_1 ,\dots,R_s\in\pi_s(\L)}\r(R_1,\s_{R_1})\cdots\r(R_s,\s_{R_s})~=~
$$
\vskip.2cm
$$
~=~ \sum_{s=1}^{|\L|}\sum_{R_1 ,\dots ,R_s\in\pi_s(\L)}
\tilde\r(R_1) \cdots
\tilde\r(R_s)~~~~~~~~~~~~~~~~~~~~~~~~~~~~~~~~~~~
$$

\\where

$$
{\tilde \r}(R)\,~=~\, \sum_{\s_{R}\in \Si_R}\;\r(R)\;e^{-\b D\sum_{x\in R}\;\s^2_x}
$$

\\Now, due to assumption {\bf A.} formula \equ(hyp0), we have  that
$
\prod_{\{x,y\}\in
E_g}[e^{-\b V(x,y, \s_x,\s_y)}-1]= 0$ for any $g\in G_R$ whenever  $\s_x =0$ for some $x\in R$.
\\Hence a straightforward computation  gives

$$
{\tilde \r}(R)~=~\cases{ 1+2\sum_{k=1}^N e^{-\b Dk^2}&if $|R|~=~1$\cr \cr
\sum_{\s_R\in \tilde\Si_R}e^{-\b D\sum_{x\in R}\;\s^2_x}
\sum_{g\in
G_R}\prod_{\{x,y\}\in E_g}[e^{-\b V(x,y, \s_x,\s_y)}-1] &if $|R|\geq 2$\cr}
$$
where $\tilde\Si_R=\{\s_R\in \Si_R: \s_x\neq 0,~\forall x\in R\}$.

\\Define now, for any finite $R\subset \VU$ such that $|R|\geq 2$
$$
\z(R)~=~\left[{1\over 1+2\sum_{k=1}^N e^{-\b Dk^2}}\right]^{|R|}
\sum_{\s_R\in \tilde\Si_R}e^{-\b D\sum_{x\in R}\;\s^2_x}
\sum_{g\in
G_R}\prod_{\{x,y\}\in E_g}[e^{-\b V(x,y, \s_x,\s_y)}-1]\Eq(defact)
$$
Then it is not difficult to check that
$$
Z_{\L}(\b) ~=~  (1+2\sum_{k=1}^N e^{-\b Dk^2})^{|\L|}~~\Xi_{\L}(\b)\Eq(polex)
$$

\\where
$\Xi_{\L}(\b)$ is given by

$$
\Xi_{\L}(\b )~=~1+\sum_{s\geq 1}{1\over n!} \sum_{R_1 ,\dots
,R_n\atop |R_i|\geq 2,\, R_i\cap R_j~=~\emptyset}
\z(R_1)\cdots\z(R_n)\Eq(gcf)
$$

\\Hence the free energy of the system is given by
$$
f_\L(\b)=\log (1+2\sum_{k=1}^N e^{-\b Dk^2})+P_\L(\b)
$$
where
$$
P_\L(\b)= {1\over |\L|}\log \Xi_{\L}(\b ) \Eq(pressu)
$$
The function $\log (1+2\sum_{k=1}^N e^{-\b Dk^2})$ is analytic for all $\b\ge 0$ so to check the analyticity of
the free energy it is sufficient
to check the analyticity of the function  $P_\L(\b )$  defined in \equ(pressu).

Now we just note that $\Xi_{\L}(\b )$ is the partition function of a hard-core polymer gas
in which the polymers $R$ are finite subsets of $\VU$ with cardinality greater than 1, activity $\z(R)$ and with
the incompatibility
relation being the non-empty intersection.
By Fernandez-Procacci criterion  \cite{FP} (see also sec. 3 in \cite{JPS}), the pressure $P_\L(\b)$
of this hard-core polymer gas is written in terms of an absolutely convergent  series  bounded uniformly in $\L$
if the activities \equ(defact) satisfy
$$
\inf_{a>0}(e^a-1)^{-1}\sum_{n\ge 2}e^{an} \sup_{x\in \VU}\sum_{R\subset \VU:\,x\in R\atop |R|~=~n} |\z(R)|\le 1 \Eq(FP)
$$
\\Now, since $e^{-\b D\sum_{x\in R}\;\s^2_x}\le e^{-\b D|R|}$ for any $\s_R\in \tilde \Sigma_R$
and setting
$$
{\tilde \l_\b} ~=~ {e^{-\b D}\over 1+2\sum_{k=1}^Ne^{-\b Dk^2}}\Eq(lti)
$$
we have
\def\lt{{\tilde \l}}
\vskip.5cm
$$
\sup_{x\in \VU}\sum_{R\subset \VU:\,x\in R\atop |R|~=~n} |\z(R)|~\le~
{\lt_\b}^n\sup_{x\in \VU}\sum_{R\subset \VU:\,x\in R\atop |R|~=~n}\sum_{\s_R\in \tilde \Si_R}
\bigg|\sum_{g\in G_R}\prod_{\{x,y\}\in E_g}[e^{-\b V(x,y, \s_x,\s_y)}-1]\bigg|\Eq(xizz)
$$
We now need to bound the factor
$$
\Bigg|\sum_{g\in
G_R}\prod_{\{x,y\}\in E_g}[e^{-\b V(x,y, \s_x,\s_y)}-1]\bigg| \Eq(factor)
$$
in the r.h.s. of the equation above.  Usually
factors like \equ(factor) are bounded using sophisticated tools in cluster expansion  such as
the so called  Battle-Brydges-Federbush tree inequality (see e.g \cite{Br}, \cite{PdLS})
or the Brydges-Kennedy-Abdesselam-Rivasseau formula (see e.g \cite{AR1}, \cite{BK}). However,
for the case considered here (bounded spins in a discrete set),  it is possible to use a much simpler  result
obtained recently in \cite{JPS} whose very simple proof
is based on the Penrose identity \cite{Pe} (see also \cite{Pf}, \cite{S1}, and \cite{FP}).
Such a result, stated in \cite{JPS} as Proposition 4.3, in the present notations can be rephrased as follows.

\\{\bf Proposition 2 (Jackson-Procacci-Sokal).}
{\it Let $P_2(\VU)$ be the set of unordered pairs of distinct elements of $\VU$ and let $\bm v:
P_2(\VU)\to \mathbb{C}: \{x,y\}\mapsto v_{\{x,y\}}$ be a
 map with the following property: there exists a (nonnegative) constant $B$ such that
$$
\sup_{x\in \VU}\sum_{y\in \VU\atop y\neq x} |v_{\{x,y\}}|\le 2B \Eq(stabil)
$$

\\Then, for
any finite $R\subset S$ with $|R|\ge 2$ the following inequality
holds:

$$
\Big|\sum_{g\in {  G}_{R}}\prod_{\{x,y\}\in E_g}
\left(e^{-v_{\{x,y\}}}-1\right)\Big|\leq
e^{B|R|}\sum_{\t\in T_R}\prod_{\{x,y\}\in E_\t}
\Big(1-e^{-|v_{\{x,y\}}|}\Big)
$$
where $T_R$ is the set of all trees   with vertex set $R$ and $E_\t$
denotes the edge set of a tree $\t\in T_R$.}
\vv
\\{\it Proof}.  Choose a root $x_0$ in $R$  so that  any tree $\t\in T_R$
is regarded as rooted in $x_0$.
Choose an order in $R$, i.e. to any $x$ in $R$ associate in a one-to-one way a number  in $\{1,2,\dots, |R|\}$ (the label of $x$).
Then for any tree $\t\in T_R$  with edge set $E_\t$ define a graph $p(\t)$ in ${  G}_{R}$ with
edge set $E_{p(\t)}\supset E_\t$ obtained  by adding to $E_\t$ all unordered pairs in $R$  connecting each vertex $x\neq x_0$
of $\t$   to all its siblings in $\t$ (vertices of the same
generation) and to all its ``uncles"  in $\t$
(vertices of the previous
generation) with label greater than the label of the parent of $x$. Then the Penrose identity is
$$
\sum_{g\in {  G}_{R}}\prod_{\{x,y\}\in E_g}
\left(e^{-v_{\{x,y\}}}-1\right)=
\sum_{\t\in T_R}\prod_{\{x,y\}\in E_\t} \left(e^{-v_{\{x,y\}}}-1\right)e^{-\sum_{\{z,z'\}\in E_{p(\t)\backslash E_\t}}v_{\{z,z'\}}}
$$

Hence
$$
\Big|\sum_{g\in { \cal G}^c_{R}}\prod_{\{x,y\}\in E_g}
\left(e^{-v_{\{x,y\}}}-1\right)\Big|\le
\sum_{\t\in T_R}\prod_{\{x,y\}\in E_\t} \left(e^{|v_{\{x,y\}}|}-1\right)e^{\sum_{\{z,z'\}\in E_{p(\t)\backslash E_\t}}|v_{\{z,z'\}}|}=
~~~~~~~~~~~~~~$$
$$
=
\sum_{\t\in T_R}\prod_{\{x,y\}\in E_\t} (1-e^{-|v_{\{x,y\}}|})\,\,e^{\sum_{\{z,z'\}\in E_{p(\t)}}|v_{\{z,z'\}}|}
\le
\sum_{\t\in T_R}\prod_{\{x,y\}\in E_\t} (1-e^{-|v_{\{x,y\}}|})\,\,e^{\sum_{\{z,z'\}\subset R}|v_{\{z,z'\}}|}\le
$$
$$
\le
\sum_{\t\in T_R}\prod_{\{x,y\}\in E_\t} (1-e^{-|v_{\{x,y\}}|})\,\,e^{{1\over 2}|R|
\sup_{z\in R}\sum_{z'\in \VU\atop z'\neq z}|v_{\{z,z'\}}|}\le
e^{B|R|} \sum_{\t\in T_R}\prod_{\{x,y\}\in E_\t} (1-e^{-|v_{\{x,y\}}|})
$$
$\Box$
\vv
Now, for $v_{\{x,y\}}=\b V(x,y, \s_x,\s_y)$, by assumption {\bf B.}, formulas \equ(hyp2) and \equ(hyp1), the pair potential $V(x,y, \s_x,\s_y)$
satisfies \equ(stabil) with $B=J$.
Therefore, proposition 2
leads to the bound
$$
\biggl| \sum_{g\in G_R} \prod_{\{x,y\}\in E_g} \bigl[ e^{-\beta V(x,y,\sigma_x,\sigma_y)} -1 \bigr]\biggr|
\;\le\; e^{\beta J |R|}\sum_{t\in T_R} \prod_{\{x,y\}\in E_t} \bigl( 1-e^{-\beta J(x,y)}\bigr)\;.
\Eq(tgi)$$
\\Hence, pugging \equ(tgi) in \equ(xizz) and  using also that $\sum_{\s_R\in \tilde \Si_R}1=(2N)^n$, we get
$$
\sup_{x\in \VU}\sum_{R\subset \VU:\,x\in R\atop |R|~=~n} |\z(R)|
~\le~ [2N\lt_\b e^{\b J}]^n\sup_{x\in \VU}\sum_{R\subset \VU:\,x\in R\atop |R|~=~n}~
\sum_{t\in
T_R}\prod_{\{x,y\}\in E_t}\bigl( 1-e^{-\beta J(x,y)}\bigr) =
$$
$$
={[2N\lt_\b e^{\b J}]^n\over (n-1)!}\sup_{x\in \VU}\sum_{t\in
T_n}\sum_{(x_1,\dots,x_n)\in \VU^n \atop x_1=x,\,\,x_i\neq x_j }
\prod_{\{i,j\}\in E_t}\bigl( 1-e^{-\beta J(x_i,x_j)}\bigr)\Eq(reorg)
$$
where $T_n$ denotes the set of trees with vertex set $\{1,2,\dots,n\}$. It is now easy to check that
$$
\sum_{(x_1,\dots,x_n)\in \VU^n \atop x_1=x,\,\,x_i\neq x_j }
\prod_{\{i,j\}\in E_t}\bigl( 1-e^{-\beta J(x_i,x_j)}\bigr)\leq [h(\b, J)]^{n-1}, ~~~\forall t\in T_n\Eq(sum)
$$
with
$$
h(\b, J)=\sup_{x\in \VU}\sum_{y\in \VU \atop y\neq x}\bigl( 1-e^{-\beta J(x,y)}\bigr)\Eq(hbet)
$$
Note that, by  \equ(hyp2), we have
$$
h(\b, J)=\sup_{x\in \VU}
\sum_{y\in \VU \atop y\neq x}\b J(x,y) \int_0^1e^{-\beta J(x,y)t}\le \sup_{x\in \VU}\sum_{y\in \VU \atop y\neq x}\b J(x,y) \le 2\b J\Eq(crude)
$$
Using finally Cayley formula (i.e. $\sum_{t\in T_{n}}1= n^{n-2}$) for tree counting, we obtain
$$
\sup_{x\in \VU}\sum_{R:\,x\in R\atop |R|~=~n} |\z(R)|~\le~ {n^{n-2}\over (n-1)!}[h(\b, J)]^{n-1} [2N\lt_\b e^{\b J}]^n
$$

\\Hence the convergence criterion \equ(FP) is
%$$
%\sum_{n\ge 2}e^{an} {n^{n-2}\over (n-1)!}(2\b J)^{n-1} (2\lt_\b e^{\b J})^n\le e^a-1
%$$
%i.e.
$$
\inf_{a>0}(e^a-1)^{-1}\sum_{n=2}^{+\infty}e^{an} \Big[2Nh(\b, J) \lt_\b e^{\b J}\Big]^{n-1} {n^{n-1}\over n!}
\le {1\over 2N\lt_\b e^{\b J}}
\Eq(6.2)
$$

\\which, by lemma 6.1 in \cite{JPS}   is satisfied when
%$$
%{1\over 4\b J \lt_\b e^{\b J}}\ge {4\over {1\over 2\lt_\b e^{\b J}}}+ 3
%$$
%i.e. if
$$
{1\over 2Nh(\b, J)\lt_\b e^{\b J}}\ge {8N\lt_\b e^{\b J}}+ 3
$$
i.e. when
%$$
%{1\over {8[\lt_\b e^{\b J}}]^2+ 3\lt_\b e^{\b J} }\ge 4\b J
%$$
%Now,
%$$
%\lt_\b e^{\b J}= {\l_\b e^{\b J}\over (1+2\l_\b)} = e^{-(D-J)\b}\k_\b
%$$
%with
%$$
%\k_\b= {1\over (1+2\l_\b)}
%$$
%thus condition becomes
$$
{e^{(D-J)\b} F(\b)}\ge h(\b, J)  \Eq(estr)
$$
with
$$
F(\b)={1\over 2}
%{(1+2\l_\b)\over {[8\lt_\b e^{\b J}}+ 3]}=
{(1+2\sum_{k=1}^N e^{-\b Dk^2})^2
\over {[8N^2e^{-(D-J)\b} }+3N(1+2\sum_{k=1}^N e^{-\b Dk^2})]}\Eq(fb)
$$

\\Now it is easy to see there always exist two positive numbers $\b_1$ and $\b_2$ (with  $\b_1<\b_2$)
depending on $D$, $J$ and $N$, such
that the inequality \equ(estr) is in general satisfied either when $\b\le \b_1$, or when $\b\ge \b_2$ and $D>J$.

Indeed,
$F(0)={1\over 4}
{(1+2N)^2
\over 3N+14N^2}$, so that for $\b$ sufficiently small, say $\b\le \b_1$, the inequality \equ(estr) is  always satisfied, because the l.h.s. tends to $F(0)> 0$ for $\b\to 0$ regardless of the value of $D$, while
in the r.h.s. $\lim_{\b\to 0} h(\b,J)=0$,  by \equ(crude).

On the other hand, 
$F(\b)$
tends to ${1\over 12N}$ as $\b\to\infty$.  So, as soon as $D>J$,  the inequality \equ(estr)
is surely satisfied for $\b$ sufficiently large, say $\b\ge \b_2$, when the exponential on the l.h.s of
\equ(estr) beats the function $h(\b,J)$. which, by \equ(crude), is always below the linear function $2\b J$.

Finally, let  us show  that there is a critical value of $D_c$ (depending on $J$) such
that, whenever $D\ge D_c$, the inequality \equ(estr) is always satisfied for all $\b\in [0,\infty)$.
Indeed, an upper bound for $D_c$ can be easily found if  we are not looking for optimal bounds. As a matter of fact
it is easy to see that

$$
F(\b)\ge {1\over 6N +16N^2},~~~~~~~~~~\mbox{for all $\b\ge 0$}\Eq(crud)
$$
Moreover using  the bound \equ(crude)
the inequality \equ(estr) is surely satisfied if
$$
{e^{(D-J)\b}}\ge {(12N +32N^2)}\b J \Eq(estr2)
$$
which is always satisfied for any $\b\ge 0$ as soon as
$
D\ge \left(1 + {12N +32N^2\over e}\right)J
$
%As a matter of fact, the minimum of the function
%$$
%f(\b)= {e^{(D-J)\b}}- {68\over 9}\b J
%$$
%occurs when
%$$
%(D-J)e^{(D-J)\b}={68\over 9}J\Eq(starr2)
%$$
%Let $\b_c$ be the solution of \equ(starr2). Then
%$$
%f(\b_c)= {68\over 9}J\left({1\over D-J}-\b_c\right)
%$$
%and imposing
%$$
%f(\b_c)= 0
%$$
%we get
%$$
%{1\over D-J}=\b_c\Eq(cass)
%$$
%and since, by \equ(starr) $(D-J)\b_c= \ln \left[ {{68\over 9}J\over (D-J)}\right]$, the condition \equ(cass) becomes
%$$
%\ln \left[ {{68\over 9}J\over (D-J)}\right]=1
%$$
%or
%$$
%{{68\over 9}J\over (D-J)}=e
%$$
%or
%$$
%D=\left(1 + {{68\over 9e}}\right)J
%$$
%This $D$ is the value such that the function $e^{(D-J)\b}$ is tangent to the function $12\b J$.

\\So we get for the critical value $D_c$ of the crystal field the upper bound
$$
D_c\le  \left(1 + {12N +32N^2\over e}\right)J
\Eq(upper)
$$
Of course, due to the rough bounds \equ(crude) and \equ(crud),  this estimate is far from being optimal.  In particular,
the replacement in \equ(estr) of the function $h(\b, J)$ with its upper bound \equ(crude) seems to be a quite crude estimate,
especially in the large $\b$ regime.
To find with accuracy the regions of validity of the inequality \equ(estr) so as to improve the
upper bound for $D_c$, one should be able
compute explicitly function  $h(\b, J)$ which depends sensibly
on the behavior of  $J(x,y)$ and this would go beyond the
scope of the present note.

We conclude by noting that improvements on the upper bound \equ(upper)
might also  be obtained by trying  to enlarge the analyticity region in  the disordered phase (i.e. when $D>J$)
via a genuine low-temperature contour expansion for the system, i.e  an expansion
around the ground state $\sigma=0$.

\section*{Acknowledgments}

It is a pleasure to thank Aernout Van Enter for many valuable comments and suggestions.
We also thank an anonymous referee whose very pertinent remarks  helped  to improve the
contents of the paper.
A. P.  thanks the Isaac Newton Institute for Mathematical Sciences,
University of Cambridge, for generous support during the programme on
Combinatorics and Statistical Mechanics (January--June 2008),
where this work was begun.
This research was supported in part by the Conselho Nacional de Desenvolvimento Cient\`\i fico e Tecnol\'ogico
(CNPq), and by the Funda\c{c}\~ ao de Amparo \`a Pesquisa do Estado de Minas Gerais
(FAPEMIG).

\end{document}